\documentclass[a4paper,11pt]{article}
\usepackage{pos}
\usepackage{accents}

\def\attdiii{\accentset{\circ}{a}^{(3)}_{\tau\tau}} 
\def\supbaylimdimiiitattfowo{2 \times 10^{-26}}

\def\DMMW{4\times 10^{-28}}
\def\DMDomi{10^{-22}}

\title{Ultra-light Dark Matter Search with Astrophysical Neutrino Flavour}
\ShortTitle{Ultra-light Dark Matter Limits from Astrophysical Neutrino Flavour}

\author[a]{Carlos~A.~Arg\"{u}elles}
\author[b]{Kareem~Farrag}
\author*[c]{Teppei~Katori}

\affiliation[a]{Harvard University, Department of Physics and Laboratory for Particle Physics and Cosmology,\\
Cambridge, MA 02138, USA}

\affiliation[b]{Chiba University, Department of Physics and Institute for Global Prominent Research,\\
Chiba, 263-8522, Japan}

\affiliation[b]{King's College London, Department of Physics,\\
London, WC2R 2LS, UK}

\emailAdd{carguelles@g.harvard.edu}
\emailAdd{k.r.h.farrag@chiba-u.jp}
\emailAdd{teppei.katori@kcl.ac.uk}

\abstract{
The ultra-light dark matter is a new class of dark matter candidates. Unlike traditional dark matter particle candidates, the ultra-light dark matter behaves like a classical field, which saturates the entire Milky Way galaxy with a coherent oscillation. If such dark matter exists and couples with neutrinos, properties of astrophysical neutrinos propagating in the Milky Way would be modified and detectable by neutrino telescopes such as IceCube. Meantime, IceCube looked for quantum-gravity-motivated effects from the astrophysical neutrino flavour information. IceCube did not find evidence of new physics, and they set limits on neutrino - Lorentz violating field couplings. Here, we investigate if these results can be applied to limit the ultra-light dark matter couplings with neutrinos. It is found that the strong limits of neutrino-dark matter couplings in low dark matter mass regions can be obtained in this approach.
}

\ConferenceLogo{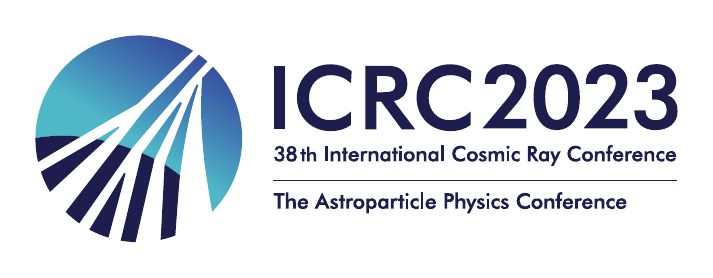}

\FullConference{%
38th International Cosmic Ray Conference (ICRC2023)\\
  26 July - 3 August, 2023\\
  Nagoya, Japan}

\begin{document}
\maketitle

\section{IceCube Lorentz violation limits}

Recently, the IceCube collaboration set the most stringent limits on Lorentz violating effects in the neutrino sector using the astrophysical neutrino flavour data~\cite{IceCube:2021tdn,Author:2023icrc}. For this, 7.5 years of the high-energy starting event (HESE) data sample is used~\cite{IceCube:2020wum}. 
Since the data statistics are limited, astrophysical neutrino flavour is represented by the flavour ratio~\cite{IceCube:2015rro,IceCube:2015gsk,IceCube:2018pgc,IceCube:2020fpi} where the fraction of electron, muon, and tau neutrinos ($\nu_e:\nu_\mu:\nu_\tau$) are fit from the data. They are understood as the sum of neutrino and anti-neutrino contributions with integrating the assumed astrophysical neutrino spectrum. Couplings between neutrinos and Lorentz violating spacetime, if existed, can deviation the flavour data from the standard astrophysical neutrino flavour prediction to anomalous value~\cite{PhysRevLett.115.161303}. The analysis used effective operators derived from the Standard-Model Extension (SME) framework~\cite{Kostelecky:2011gq} assuming all effects are isotropic. For example, under the SME, the effective Hamiltonian of astrophysical neutrinos with dimension-three Lorentz violating operators can be written explicitly in the flavour basis,
\begin{eqnarray}
H_{eff}
&\sim&
\frac{1}{2E}
\cdot
\left(\begin{array}{ccc}
m^{2}_{ee} & m^{2}_{e\mu} & m^{2}_{\tau e} \\
m^{2*}_{e\mu} & m^{2}_{\mu\mu} & m^{2}_{\mu\tau}\\
m^{2*}_{\tau e} & m^{2*}_{\mu\tau} & m^{2}_{\tau\tau}
\end{array}\right)
+\left(\begin{array}{ccc}
\accentset{\circ}{a}^{(3)}_{ee} & \accentset{\circ}{a}^{(3)}_{e\mu} & \accentset{\circ}{a}^{(3)}_{\tau e} \\
{a}^{(3)}_{e\mu} & {a}^{(3)}_{\mu\mu} & {a}^{(3)}_{\mu\tau} \\
{a}^{(3)}_{\tau e} & {a}^{(3)}_{\mu\mu} & \accentset{\circ}{a}^{(3)}_{\tau\tau}
\end{array}\right)~.
\label{eq:hamiltonian}
\end{eqnarray}
Here, the first term is the neutrino mass matrix~\cite{Esteban:2020cvm}, and the second term is the dimension-three isotropic Lorentz violating term~\cite{IceCube:2017qyp}. Although they did not find the evidence of presence of these terms in the current IceCube flavor data, they set order magnitude stronger limits on certain types of couplings. For example, the dimension-three Lorentz violating operator in $\tau-\tau$ sector is constrained down to $\attdiii<\supbaylimdimiiitattfowo$~GeV with the Bayes factor greater than 10. Interestingly, these limits can be re-interpreted to constrain other physics~\cite{Arguelles:2022tki}. In these proceedings, we investigate the flavour effect of neutrino - dark matter couplings~\cite{Berlin:2016woy,Reynoso:2016hjr,deSalas:2016svi,Rasmussen:2017ert,Farzan:2018pnk,Smirnov:2019cae,Karmakar:2020yzn,Losada:2021bxx,Gherghetta:2023myo,Huang:2018cwo,Brzeminski:2022rkf,Alonso-Alvarez:2023tii} from the IceCube Lorentz violation limits.

\section{Ultra-Light Dark Matter}

Ultra-light dark matter is a class of dark matter models where the mass of dark matter particles is as low as $\DMDomi$~eV and they can still be the dominant source of dark matter~\cite{Hu:2000ke,Marsh:2015xka}. The searches of the weakly interaction massive particles (WIMPs) use elastic scatterings between WIMPs and nuclear targets~\cite{XENON:2018voc,PandaX-4T:2021bab,LZ:2022ufs}. On the other hand, ultra-light dark matter behaves as a classical field and it may saturate in our galaxy with maintaining the coherent oscillation. In this scenario, if they interact with neutrinos, the propagation of astrophysical neutrinos in our galaxy may be affected. Although interactions are so small and it cannot affect the energy spectrum, the effect may be detectable from anomalous flavour mixings inconsistent with the standard astrophysical scenario.

To set up our analysis, we make several assumptions. First, we assume dark matter fields and astrophysical neutrino couplings are described by vector or axial vector operators~\cite{Berlin:2016woy,Reynoso:2016hjr,deSalas:2016svi,Rasmussen:2017ert,Farzan:2018pnk,Smirnov:2019cae,Karmakar:2020yzn,Losada:2021bxx,Gherghetta:2023myo}, or dark matter fields themselves are vector or axial vector fields (such as neutrinophilic axion~\cite{Huang:2018cwo}, vector dark matter~\cite{Brzeminski:2022rkf,Alonso-Alvarez:2023tii}), because these interactions make matter potential terms in vacuum and they can be identified with Lorentz violating operators in Eq.~\ref{eq:hamiltonian}.
Second, we assume a constant dark matter density, $\rho=0.3$~GeV/m$^3$, across the whole Milky Way galaxy to simplify the problem. The IceCube HESE astrophysical neutrino data sample is a diffuse sample and it contains neutrinos from all directions. The ultra-light dark matter search with neutrinos can be improved by assuming more realistic dark matter density profiles, such as NFW potential~\cite{Navarro:1995iw}.
Third, we assume the flavour conversion is adiabatic~\cite{Akhmedov:2017mcc}.
This condition is energy-dependent and it is broken for fast oscillating dark matter field.
Under these conditions, we can write down the dimension-three $\tau-\tau$ sector isotropic SME coefficient in terms of the neutrino - dark matter coupling. By using the dark matter with mass $m_a$ and coupling with tau neutrino $g_\tau$,

\begin{eqnarray}
 \attdiii=g_{\tau}\frac{\sqrt{2\rho}}{m_a}\cos(m_at)~.
 \label{eq:uldm}
\end{eqnarray}

And fourth, we assume the oscillation of the ultra-light dark matter field is fast enough compared with the time scale for neutrinos to cross the Milky Way. 
The mass of the dark matter with the de Broglie length of the radius of the Milky Way is around $\DMMW$~eV. Although such mass is not forbidden as a candidate of the galactic dark matter~\cite{Marsh:2015xka}, we focus on the region greater than $\DMDomi$~eV where the ultra-light dark matter can be the dominant source of the galactic dark matter. In this scenario, astrophysical neutrinos crossing the Milky Way experience many oscillations of the dark matter field. Furthermore, neutrinos propagate different distances in the Milky Way, and the phase of the oscillation is different for neutrinos detected by IceCube. The combination of these effects makes the couplings between neutrinos and dark matter fields to be averaged out, and the expectation value becomes zero.
However, interactions of neutrinos and the oscillating dark matter fields spread the astrophysical neutrino flavour at arrival and they are still detectable~\cite{Losada:2021bxx,Hamaide:2022rwi}.
Now, we investigate this effect on the flavour triangle of the IceCube astrophysical neutrino flavour.

\section{Flavor ratio from neutrino - dark matter coupling}

\begin{figure}[t!]
 \begin{center}
 \includegraphics[width=0.6\columnwidth]{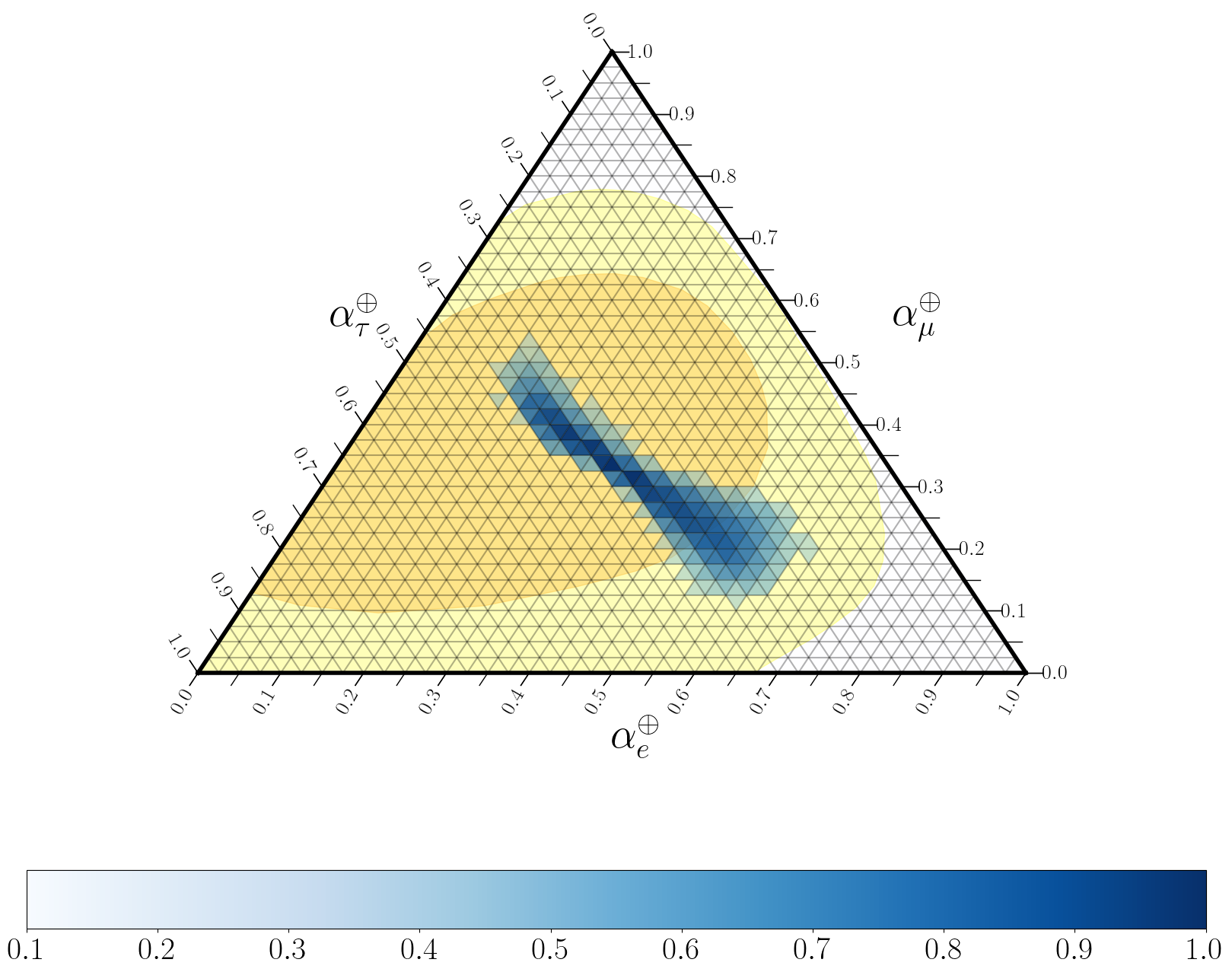}
 \end{center}
\vspace{-2mm}
\caption{Astrophysical neutrino flavour without new physics. Each point in this figure is the flavour ratio of astrophysical neutrinos at detection. The histogram is the expected flavour ratio for the standard neutrino model with the standard astrophysical neutrino production scenarios, and it is enclosed by the $68\%$ and $95\%$ contours from the IceCube data~\cite{IceCube:2020fpi}.}
\label{fig:1}
\end{figure}

Fig.~\ref{fig:1} represent the flavour triangle of the IceCube HESE sample~\cite{IceCube:2020wum,IceCube:2020fpi}. Each point in this space shows the fraction of electron, muon, and tau neutrinos ($\nu_e:\nu_\mu:\nu_\tau$). A ``bowtie'' shape at the central region around $(\nu_e:\nu_\mu:\nu_\tau)\sim (1:1:1)$ is the expected flavour ratio from the standard astrophysical neutrino scenario where flavour ratios at the productions are all possible combinations of electron neutrinos and muon neutrinos, or $(\nu_e:\nu_\mu:\nu_\tau)\sim (x:1-x:0)$ with $x=[0,1]$. In other words, the standard astrophysical neutrino production scenario corresponds to the right side of this triangle plot. Then, the neutrino standard model Hamiltonian, Eq.~\ref{eq:hamiltonian} without the second term,
will bring them to this bowtie region by the standard flavour mixings. If the data flavour ratio is outside of this region, there is a chance that is due to new physics.

The shaded regions represent the $68\%$ and $95\%$ data contours from the IceCube HESE 7.5-yr data analysis~\cite{IceCube:2020wum,IceCube:2020fpi}. These contours enclose this astrophysical standard scenario phase space completely, meaning we would not find any new physics which deviates slightly from there even if it exist. Therefore, future improved analysis, larger statistics and better flavour sensitivity, will be essential to look for new physics through the astrophysical neutrino flavour~\cite{Arguelles:2022tki}.

\begin{figure}[t!]
 \begin{center}
 \includegraphics[width=0.6\columnwidth]{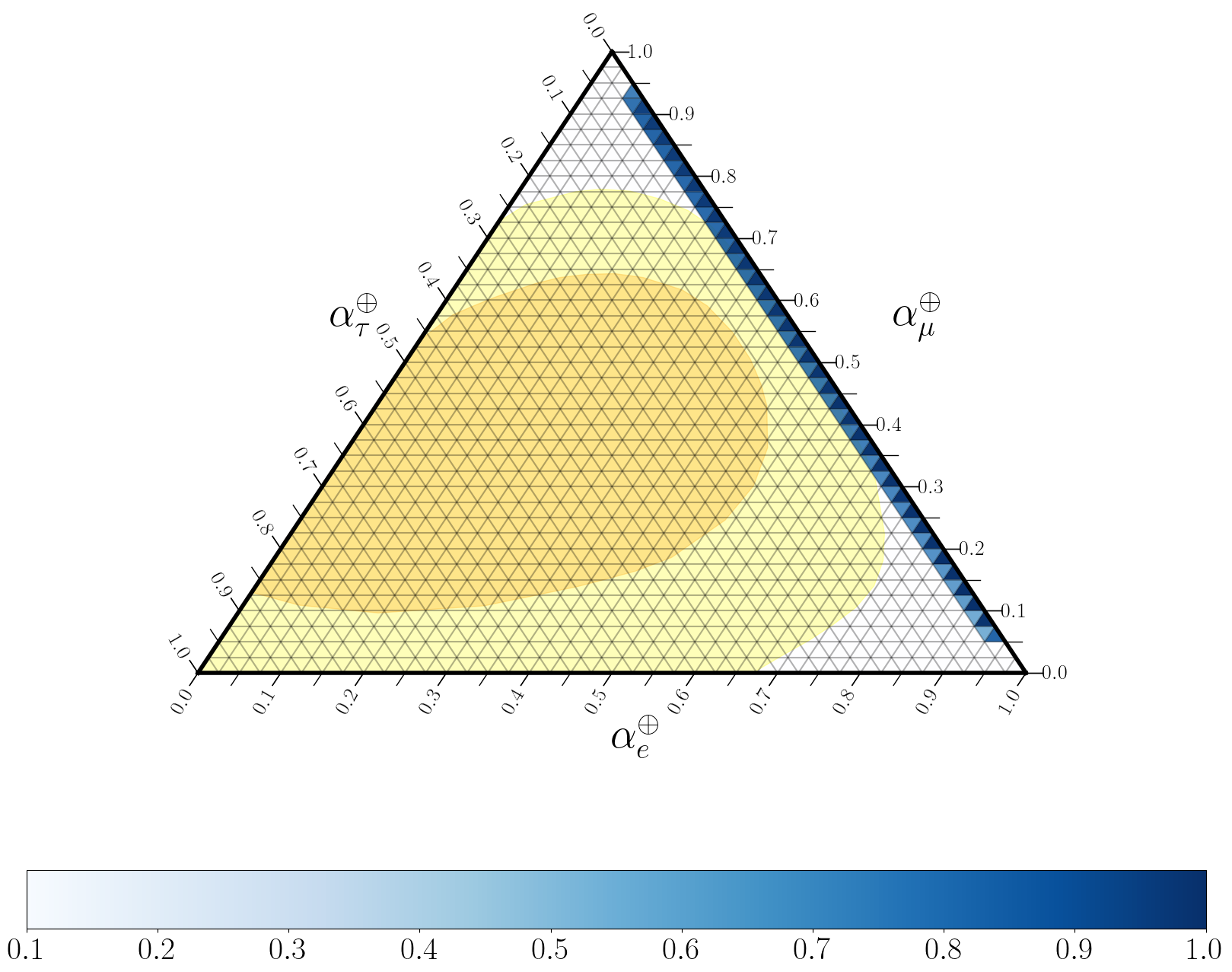}
 \end{center}
\vspace{-2mm}
\caption{Astrophysical neutrino flavour with static dark matter potential, $g_{\tau}\sqrt{2\rho}/m_a=\supbaylimdimiiitattfowo$~GeV. The value of dark matter is chosen because this value was rejected from the IceCube Lorentz violation analysis~\cite{IceCube:2021tdn}. Unlike Fig.~\ref{fig:1}, predicted astrophysical neutrino flavour is mostly outside of the $68\%$ and $95\%$ data contours~\cite{IceCube:2020fpi}.}
\label{fig:2}
\end{figure}

By adding the second term of Eq.~\ref{eq:hamiltonian}, the predicted flavour distribution is drastically changed. Fig.~\ref{fig:2} includes nonzero $\attdiii$ matrix element with the size of the IceCube limit ($\supbaylimdimiiitattfowo$~GeV) to see the effect in the flavour triangle plot. The predicted flavour distribution changes from a bowtie shape in Fig.~\ref{fig:1} to a line on the right side, and most of the points go to the outside of the data contours. This is consistent that a precise analysis found that this scenario is rejected in the IceCube with Bayes factor $>$ 10.

In terms of ultra-light dark matter, Eq.~\ref{eq:uldm}, Fig.~\ref{fig:2} corresponds to the presence of neutrino - dark matter coupling, with the dark matter amplitude $g_{\tau}\sqrt{2\rho}/m_a=\supbaylimdimiiitattfowo$~GeV but no time-varying oscillation. The interpretation of the IceCube data is that such dark matter model is rejected, therefore, we set the limit of the combination of the neutrino - dark matter coupling and the dark matter mass,

\begin{eqnarray}
 \frac{g_{\tau}}{m_a}< 10^{-14}~eV^{-1}~.
 \label{eq:limit} 
\end{eqnarray}

This limit is stronger than the current and future limits from accelerator-based, atmospheric, and solar neutrino experiments~\cite{Brzeminski:2022rkf,Alonso-Alvarez:2023tii}. This is because the expected effect scales with energy, and the high-energy astrophysical neutrinos have higher energy than other known neutrinos and strongly constrain neutrino - dark matter couplings. However, such limit may not be applicable if the ultra-light dark matter field is time varying within our galaxy.

\begin{figure}[t!]
 \begin{center}
 \includegraphics[width=0.6\columnwidth]{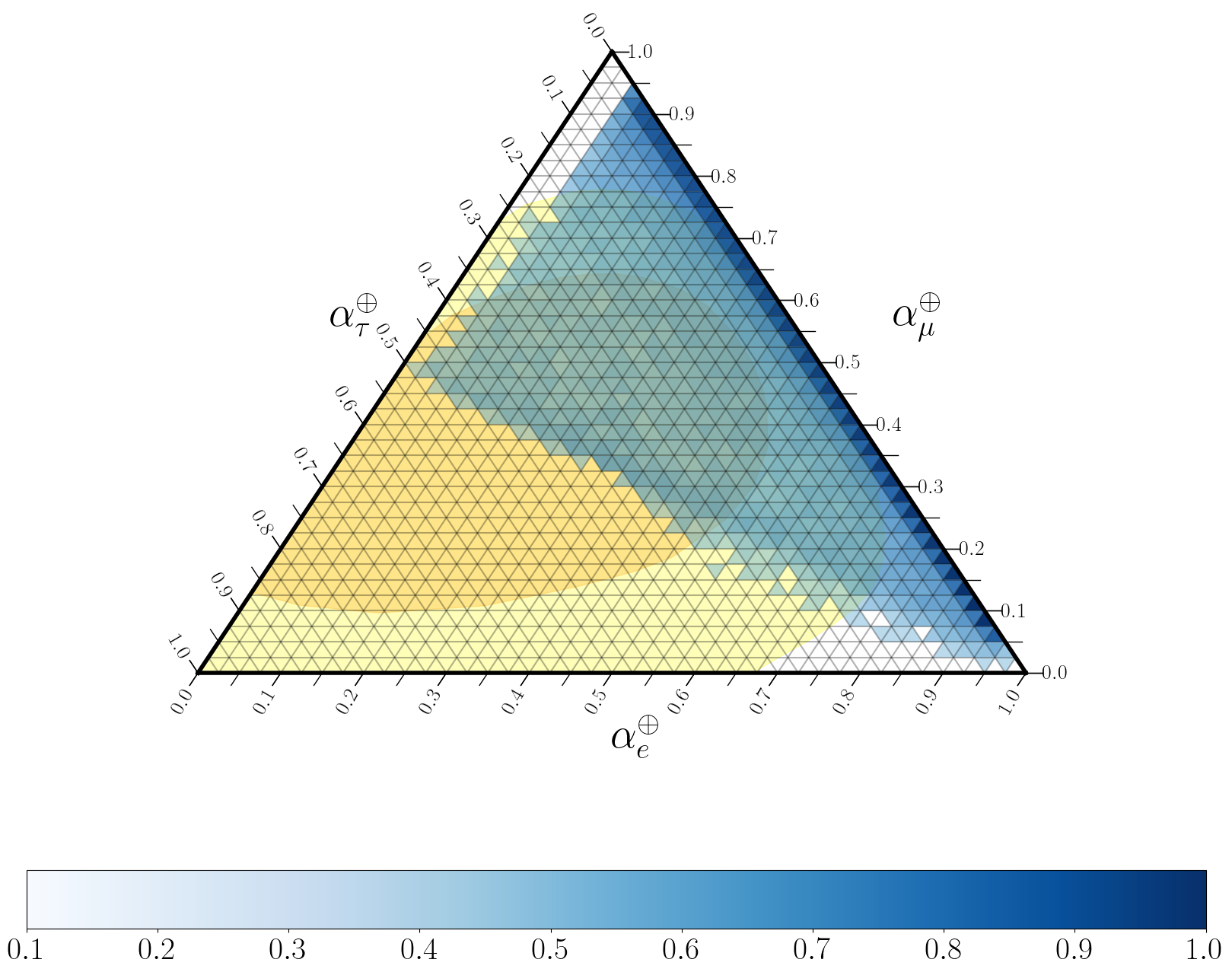}
 \end{center}
\vspace{-2mm}
\caption{Astrophysical neutrino flavour with oscillating dark matter potential, Eq.~3, with $g_{\tau}\sqrt{2\rho}/m_a=\supbaylimdimiiitattfowo$~GeV. The value of the dark matter field amplitude is chosen from the IceCube Lorentz violation analysis~\cite{IceCube:2021tdn}. Unlike Fig.~\ref{fig:2}, some of predicted astrophysical neutrino flavour predictions are enclosed by the $68\%$ and $95\%$ data contours~\cite{IceCube:2020fpi}.
}
\label{fig:3}
\end{figure}

Fig~\ref{fig:3} shows the flavour ratio with Eq.~\ref{eq:hamiltonian}, but here nonzero $\attdiii$ is replaced with Eq.~\ref{eq:uldm} to simulate the time-dependent neutrino - dark matter coupling. If we include the time variation of the dark matter field, since the potential is zero on average, one would expect that the dark matter potential will not affect the astrophysical neutrino flavour ratio. However, the time variation of the field adds the smearing feature on the flavour triangle~\cite{Losada:2021bxx,Hamaide:2022rwi}, and it spreads the predicted flavour ratio distribution in this triangle. The difference between Fig.~\ref{fig:2} and Fig.~\ref{fig:3} means these two models receive different constraints from the astrophysical flavour data. In general, we expect a weaker limit for the oscillating dark matter models than the static ultra-light dark matter models.

\section{Outlook}

Here, we briefly present a new way to look for ultra-light dark matter from the IceCube astrophysical neutrino flavour. In particular, we investigate the way to translate the recent IceCube Lorentz violation limits~\cite{IceCube:2021tdn,Author:2023icrc} to the limits of neutrino - ultra-light dark matter couplings. The expected limits from this analysis are likely to be very strong, and future neutrino telescopes, such as IceCube-Gen2~\cite{IceCube-Gen2:2020qha,Gen2:2023icrc}, can perform an improved ultra-light dark matter search with astrophysical neutrino flavour.

\section*{Acknowledgements}
CAA is supported by the Faculty of Arts and Sciences of Harvard University, and the Alfred P. Sloan
Foundation, USA. KF is supported by the JSPS and KAKENHI, Japan. TK is supported by UKRI STFC, UK. 
We thank the organizers of the ICRC2023 conference for hosting this largest-ever ICRC conference with great fun.

\bibliographystyle{ICRC}
\bibliography{ULDM}

%

%
%
%

\end{document}